# Large-Area Superconducting Nanowire Single-Photon Detector with Double-Stage Avalanche Structure

Risheng Cheng, Menno Poot, Xiang Guo, Linran Fan and Hong X. Tang

*Abstract*—We propose a novel design of superconducting nanowire avalanche photodetectors (SNAPs), which combines the advantages of multi-stage avalanche SNAPs to lower the avalanche current $I_{AV}$ and that of series-SNAPs to reduce the reset time. As proof of principle, we fabricated 800 devices with large detection area (15 μm × 15 μm) and five different designs on a single silicon chip for comparison, which include standard SNSPDs, series-3-SNAPs and our modified series-SNAPs with double-stage avalanche structure 2*2-SNAPs, 2*3-SNAPs, and 3*3-SNAPs. The former three types of the detectors demonstrate fully-saturated device detection efficiencies of ~20% while the latter two types are latching at larger bias currents. In addition, the $I_{AV}$ of 2*2-SNAPs is only 64% of the switching current $I_{SW}$ that is lower than series-3-SNAPs (74%) and well below that of 4-SNAPs (84%) reported elsewhere. We also measure that the exponential decay times of the detectors are proportional to $1/n^2$ due to the lack of external choke inductors. In particular, the decay time of 3*3-SNAPs is only 0.89 ns compared to the standard SNSPDs' 63.2 ns, showing the potential to attain GHz counting rates.

## I. Introduction

Superconducting nanowire avalanche photodetectors with $n$ parallel nanowires ($n$-SNAPs) [1], [2], also known as cascade-switch superconducting nanowire single-photon detectors (CS-SNSPDs) [3], [4], have shown several advantages over standard SNSPDs with single nanowire element, including an $n$ times improved signal-to-noise ratio (SNR) and thus reduced timing jitter. However, the avalanche current $I_{AV}$ – the minimum bias current that can trigger the avalanche switching – increases significantly with $n$ and rapidly approaches the switching current $I_{SW}$, narrowing the bias window. In practice, $n$ is limited to four when requiring saturated detection efficiency because of the inevitable inhomogeneity between individual nanowires [1]. Zhao et al. [5] implemented 8-SNAPs with multiple-avalanche architecture to reduce the $I_{AV}$ and demonstrated eight-fold signal amplification, whereas the reset time greatly increased due to the choke inductors arranged in a binary-tree layout. On the other hand, series-SNAP structure has proven to be very effective in shortening the reset time in our previous work [6] and work by Murphy et al. [7].

Here, we present a novel design of SNAPs combining the advantages of successive avalanches and series-SNAPs. Using double-stage successive avalanche structure, the $I_{AV}$ can be considerably reduced for the same total number of parallel nanowires and hence enable $n > 4$, while maintaining the merit of the $n$ times improved SNR. Meanwhile, by adopting the series-SNAP structure, the choke inductors, which are key to the current redistribution, are entirely folded into the active detection area, and thus the reset time can also be significantly reduced by $1/n^2$. Due to the improved SNR, double-stage series-SNAPs also present better timing jitter than standard SNSPDs and are comparable with previous single-stage SNAPs.

## II. Device Design and Fabrication

Fig. 1(a) shows an equivalent electrical circuit model for the series-3-SNAP. A series-$n$-SNAP consists of several serially connected $n$-SNAPs with $n$ parallel nanowires, each of which is modeled as an inductor $L_0$. When a single photon is absorbed by one of the nanowire elements and thus create a hotspot, the bias current has to be diverted into the neighboring parallel nanowires first because of the current-limiting choke inductor $L_s$, which is much larger than $L_0$ in large-area detectors. If the bias current $I_{bias}$ is high enough and larger than the avalanche current $I_{AV}$, the diverted current switches the secondary ($n$-1) nanowire elements to the normal state. Therefore, most of the current flowing through the device, which is about $n$ times the current carried by a single nanowire element, is finally output to the load resistance $R_{out}$, providing $n$ times signal-to-noise ratio (SNR) compared to standard SNSPDs. In previous $n$-SNAP structures, the series choke inductor $L_s$ is connected externally and typically designed as 10 times larger than $L_0$ for ensuring stable operation without after-pulsing [8], which limits the reset time and also increases the timing jitter due to slowed rising edge of output pulses. However, in this modified series-$n$-SNAP structure, all the unfired $n$-SNAPs serve as choke inductor $L_s$ until the avalanche happens. If the active detection area of the detector is large and the total length of the nanowire is long enough, a dedicated external inductor is no longer needed and hence the reset time can be shortened considerably.

Fig. 1(b) illustrates the electrical schematic for our new design double-stage series-SNAP that we refer to as $k*m$-SNAP. In this illustration we set $k = 2$ and $m = 3$ as example.

The authors are with the Department of Electrical Engineering, Yale University, New Haven, CT 06511, USA (e-mail: risheng.cheng@yale.edu.)

The $k*m$-SNAP consists of a number of basic blocks connected in series and folded into the meander shape in practice, each of which is $k$ of series-$m$-SNAPs connected in parallel. The series-$m$-SNAP could further be disassembled into several serially connected $m$-SNAPs with $m$ parallel nanowires bundled by nano-bridges as shown in Fig. 2. Once one of nanowire element is fired by an absorbed photon, it triggers its ($m$-1) neighbors first just as done in series-$m$-SNAPs with the help of the choke inductor $L_{s1}$, and then the $m$ nanowires as a group continue to switch the remaining ($k$-1) × $m$ nanowires to the normal state at the second stage of the avalanche. Eventually, $k \times m = n$ times the current carried by a single nanowire element is diverted into the load resistance $R_{out}$, providing an output signal which is $n$ times that of standard SNSPDs. By adopting the series-SNAP structure, the choke inductors $L_{s1}$ and $L_{s2}$, key to the first and the second stage of avalanches, are entirely folded into the active detection area, and thus we expect the reset time can be significantly reduced by a factor of $1/n^2$. Due to the double-stage avalanche structure, we also expect the $I_{AV}$ of $k*m$-SNAPs ($k \times m = n$) could be significantly lowered in comparison with single-stage series-$n$-SNAPs with the same total number of parallel nanowires.

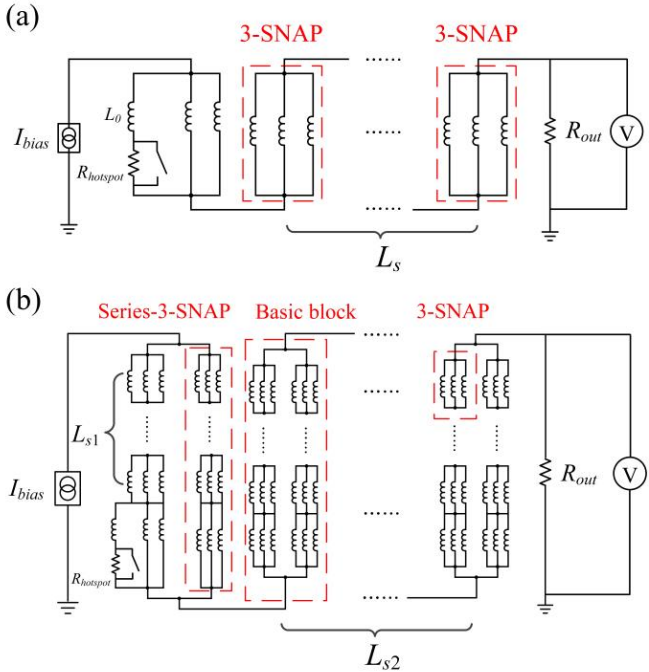

Fig. 1. Equivalent electrical circuit diagram for the series-3-SNAP (a) and the 2*3-SNAP (b). Every nanowire element is modelled as an inductor $L_0$. $I_{bias}$ represents the bias current and $R_{out}$ the load resistance. (a) The series-3-SNAP consists of a number of serially connected $n$-SNAPs. At the avalanche stage all the unfired 3-SNAPs serve as the current-limiting choke inductor $L_s$ instead of dedicated external series inductor in conventional $n$-SNAPs. (b) The 2*3-SNAP consists of a number of basic blocks connected in series and folded into the meander shape, each of which is double series-3-SNAPs connected in parallel. The series-3-SNAP could further be disassembled into several serially connected 3-SNAPs with three parallel nanowires bundled by nano-bridges. At the first stage of the avalanche the fired nanowire triggers its two neighbors first just as done in series-3-SNAPs with the help of the choke inductor $L_{s1}$. Then, the three nanowires as a group continue to switch their neighbor another series-3-SNAP to the normal state at the second stage of the avalanche, while all the remaining nanowires play the role of the choke inductor $L_{s2}$.

As proof of principle, we fabricated 800 NbTiN detectors with five different designs on the same silicon chip (with 200 nm $Si_3N_4$ on top), including standard SNSPDs, 3-SNAPs, 2*2-SNAPs, 2*3-SNAPs, and 3*3-SNAPs. Fig. 2 shows scanning electron micrographs (SEMs) of a 3*3-SNAP. The thickness of the NbTiN film, nanowire width, pitch and the dimension of the active nanowire area is 6.5nm, 38 nm, 140 nm, and 15 μm × 15 μm, respectively. Apart from different electrical structures, all the detectors are fabricated with the same design parameters for direct performance comparison. The floating nanowires surrounding the active detection area are for proximity effect correction during the e-beam exposure. Details of the devices fabrication can be found in [6].

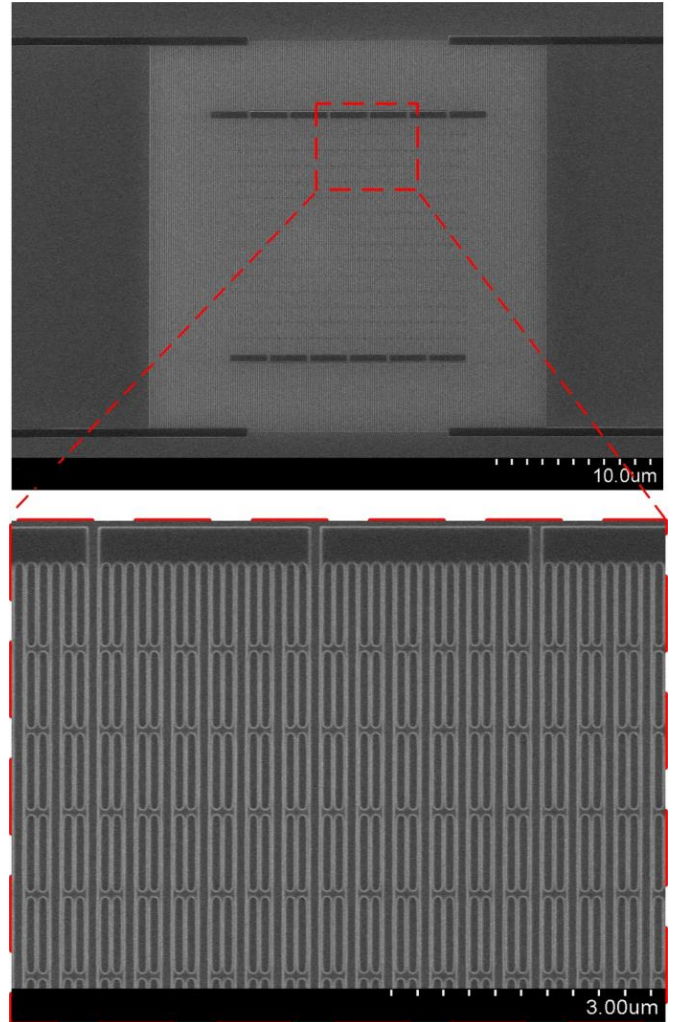

Fig. 2. Scanning electron micrographs (SEMs) of a 3*3-SNAP with higher magnification image taken at the edge of the active nanowire area. The detector consists of 6.5 nm-thick and 38 nm-wide NbTiN nanowires with 140 nm pitch and 15 μm × 15 μm active detection area. The basic nanowire elements are bundled three by three using nano-bridges to form a 3-SNAP and six of the series-3-SNAPs with a total number of 18 parallel nanowires are connected at U-turn corners together. Two of the four leads are dummy, which are placed just for symmetry.

### III. DEVICE CHARACTERIZATION

The detector chip is mounted on a 3-axis stack of Attocube stages inside a closed-cycle refrigeration cryostat [9] and cooled down to 1.7 K temperature. 1550 nm continuous wave (CW) laser light is attenuated to the single-photon level and sent to the sample chip via a standard telecommunication fiber installed in the cryostat. The detectors are flood-illuminated by fixing the fiber tip at the distance of 10 mm from the surface of the sample chip. The diameter of the provided beam spot on the chip is estimated to be around 2 mm based on the fiber-to-chip distance and the numerical aperture of the fiber. We control the Attocube stages by the LabVIEW program to move the sample chip and make the electrical contact between the RF probes and the gold pads of the detectors. All the detectors were measured and screened by an automated measurement program, and several best detectors with highest switching currents were selected for more detailed performance characterization afterwards.

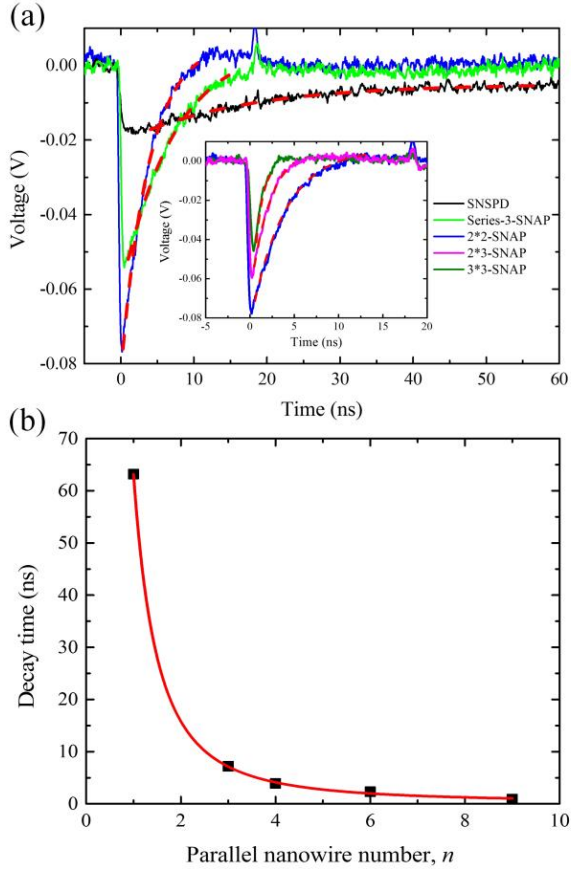

Fig. 3. (a) Single-shot traces of output pulses from five different detectors measured by a 6 GHz oscilloscope through a low-noise amplifier. The pulse heights of the 2*2-SNAP and the series-3-SNAP are approximately 4 and 3 times that of the standard SNSPD. The decay times extracted from exponential fitting for the standard SNSPD, series-3-SNAP, 2*2-SNAP, 2*3-SNAP, and 3*3-SNAP are 63.2 ns, 7.9 ns, 3.9 ns, 2.3 ns, and 0.89 ns, respectively. The former three detectors are biased at 80% of their switching currents $I_{SW}$, while the latter two detectors are biased just below the latching current (32 µA) to avoid latching. (b) Fitting of the detectors' decay times by the equation $\tau = A/n^x + b$. The best fitting is done by the set of parameters $A = 62.9$ ns, $b = 0.30$ ns, and $x = 2.02$.

Fig. 3(a) shows single-shot traces of output pulses from five different detectors amplified by a low-noise amplifier (LNA-1450, 0.01-1450 MHz bandwidth) and measured by a 6 GHz oscilloscope. We intentionally choose an amplifier with small lower cut-off frequency to avoid damped oscillation of the circuits. As expected, the pulse heights of the 2*2-SNAP and the series-3-SNAP are approximately 4 and 3 times that of the standard SNSPD, respectively. The decay times extracted from the exponential fitting for the 2*2-SNAP and the series-3-SNAP are 3.9 ns and 7.9 ns, which are respectively 1/16 and 1/9 of the standard SNSPD's 63.2 ns. The bias currents of each detectors are set to 80% of their own switching currents (10 µA, 30 µA, and 40 µA). We find all the 2*2-SNAPs, 2*3-SNAPs, and 3*3-SNAPs latching at the bias currents higher than 32 µA. The inset of Fig. 3(a) shows output pulses measured for these three types of detectors when biased just below the latching current. The exponential decay times of the 2*3-SNAP and the 3*3-SNAP are 2.3 ns and 0.89 ns, respectively. As the returning currents of the detectors are also proportional to $n$, the 2*3-SNAP and the 3*3-SNAP show relatively lower peaks compared to the 2*2-SNAP when they are all biased at the same currents. As demonstrated in Fig. 3(b), the decay times are well fitted by the equation $\tau = A/n^x + b$, where $n$ is the total number of parallel nanowires. The best fitting is done by the set of parameters $A = 62.9$ ns, $b = 0.30$ ns, and $x = 2.02$, which indicates that the reset time of the detectors with a total of $n$ parallel nanowires is reduced by $1/n^2$ since we do not have to rely on external choke inductors.

Fig. 4(a) shows the plot of photon and dark counting rates as a function of the normalized bias current $I_{bias}/I_{SW}$ for the 2*2-SNAP, series-3-SNAP, and the standard SNSPD. We can clearly see one inflection point on the 2*2-SNAP curve at 64% normalized bias current, which we define as the avalanche current $I_{AV}$. This is just slightly higher than 61% of single-stage series-2-SNAPs reported in our previous work [6] and well below that of 4-SNAPs (84%) [10] with the same number of parallel nanowires. In comparison, there are two inflection points on the curve of series-3-SNAP and we define $I_{AV}$ as 74% of $I_{SW}$ from the higher inflection point. Fig. 4(b) is the linear-scale plot of normalized counting rates depending on $I_{bias}/I_{SW}$. All the counting rates are rescaled to the maximum counting rates of the standard SNSPD achieved at the highest bias current without jump. All the three curves show distinct saturation at higher bias region and overlap nicely at the bias region higher than $I_{AV}$, indicating the single-photon detection regime of the SNAP devices. As mentioned above, we could only bias the 2*2-SNAP to 32 µA due to the latching, which is 80% of its $I_{SW}$ of 40 µA. However, the efficiency of the detector is already saturated at this point and the counting rates is 97% of the maximum value. Based on the active detector areas, the diameter of the beam-spot and calibrated power of the incident light, we roughly estimate the device detection efficiencies of the detectors are all saturated at ~20%, limited by the finite photon absorption rates without photon-cycling cavities.

We have excellent control of the fabrication process and also the quality of the film, which guarantees a decent fabrication yield. Fig. 4(c) illustrates the histogram for average switching currents of single nanowire elements $I_{SW}/n$. Despite the large areas of the detectors, more than 80% of them show $I_{SW}/n > 7.5$ µA, which indicates that these detectors could reach above 90% of the efficiencies of the best detectors as shown in Fig. 4(b).

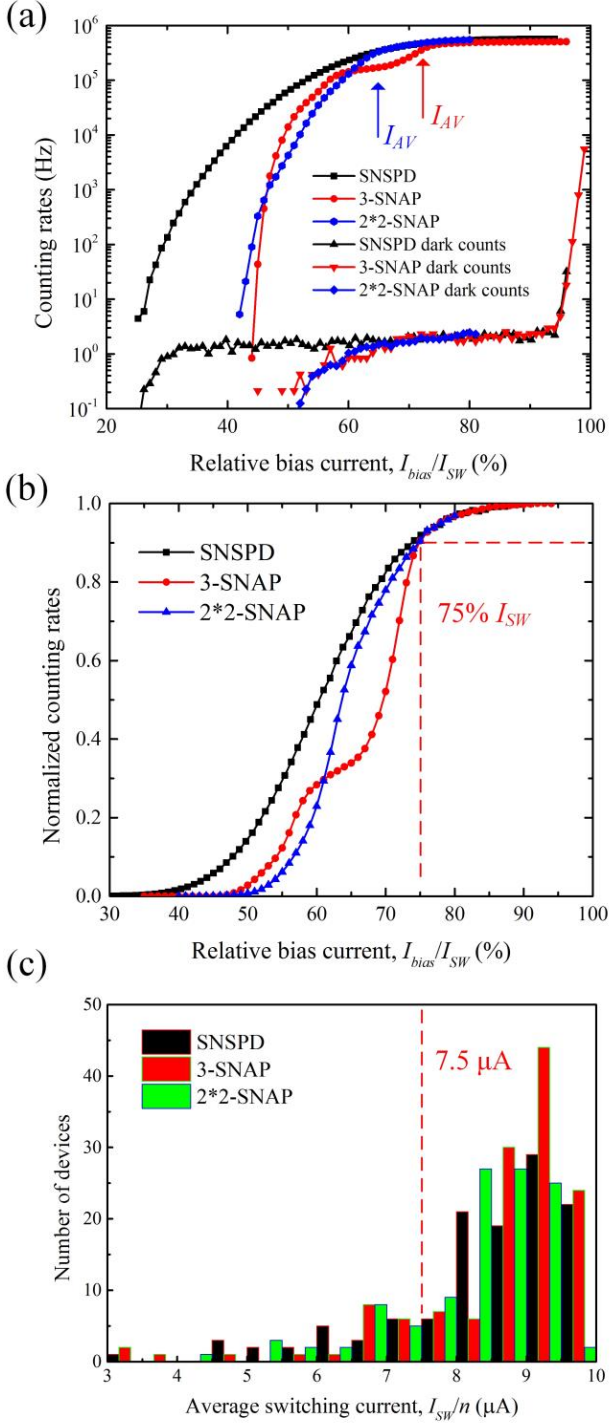

Fig. 4. (a) Photon and dark counting rates as a function of the normalized bias current $I_{bias}/I_{SW}$ for the standard SNSPD, the series-3-SNAP, and the 2*2-SNAP. The blue arrow denotes the avalanche current $I_{AV}$ of the 2*2-SNAP at 64% $I_{SW}$, while the red one for $I_{AV}$ of the series3-SNAP at 74% $I_{SW}$. (b) Normalized counting rates as a function of $I_{bias}/I_{SW}$. Perfect overlap between different detector curves at bias region $I_{bias}>I_{AV}$ indicates that the SNAP devices operate at single-photon detection regime. The two red dashed lines represent that the efficiencies of all the three detectors reach 90% of their maximum when they are biased at 75% $I_{SW}$. (c) Histogram of average switching currents of single nanowire elements $I_{SW}/n$ for different detector designs. More than 80% of the detectors show $I_{SW}/n > 7.5$ μA, corresponding to > 90% of the highest efficiencies of the best detectors.

We also measured the jitter of the detectors by sending the synchronization signal from a 4 ps-pulsed laser as well as output pulses of detectors amplified by two stages of low-noise amplifiers (LNA-1450, 0.01-1450 MHz bandwidth) to a 6 GHz oscilloscope. The jitter of the 2*2-SNAP and the series-3-SNAP is measured to be 41.9 ps and 39.9 ps respectively compared to 55.6 ps of the standard SNSPD. All values are full width at half maximum (FWHM) and measured with all the detectors biased at 80% $I_{SW}$. These improvements are attributed to the increased SNR. The small difference between the 2*2-SNAP and the series-3-SNAP indicates that the double-stage avalanche process does not induce significant extra jitter. One could expect further improvement of the timing performance by the use of cryogenic amplifiers to minimize the contribution of the jitter from the readout noise.

## IV. CONCLUSION

In conclusion, we have demonstrated a modified design of series-SNAPs with double-stage successive avalanche structure. The avalanche current $I_{AV}$ of the detectors is remarkably lowered down in comparison with conventional SNAPs with the same number of parallel nanowires $n$, i.e. the detectors can operate at lower bias current and thus lower dark counts. By folding all the choke inductors into the active detection area, the reset time of the detectors could be significantly reduced by $1/n^2$ compared to standard SNSPDs, while the SNR is improved by $n$ times. In particular, the exponential decay time of the 3*3-SNAP is only 0.89 ns compared to the standard SNSPDs' 63.2 ns, showing the potential to attain GHz counting rates. In order to solve the latching problem at higher bias current, thinner films and narrower nanowires could be fabricated to reduce the whole switching current. In addition, this new design is particularly useful for detectors made of fast-emerging low-$T_c$ amorphous superconducting materials, such as MoSi [12] or WSi [13], which have lower critical currents at 2 K but more uniform than NbN or NbTiN. Furthermore, three-stage avalanche series-SNAPs with the dimension of hundreds of micrometers could be fabricated without sacrificing the maximum counting rates.

## ACKNOWLEDGMENT

The authors would like to thank Michael Power, James Agresta, Christopher Tillinghast, and Dr. Michael Rooks for their assistance provided in devices fabrication. The fabrication of the devices was done at the Yale School of Engineering & Applied Science (SEAS) Cleanroom and the Yale Institute for Nanoscience and Quantum Engineering (YINQE).